\documentclass[]{emulateapj}

\bibpunct{(}{)}{;}{a}{}{,}
\usepackage{hyperref}
\hypersetup{
    colorlinks,
    citecolor=blue,
    filecolor=blue,
    linkcolor=blue,
    urlcolor=blue
}
\slugcomment{Accepted to ApJ} 
\shorttitle{Varied Cosmologies $\&$ Halo Substructure}
\shortauthors{Dooley et al.}

\begin{document}
\title{The Effects of Varying Cosmological Parameters on Halo Substructure}
\author{Gregory A. Dooley\altaffilmark{1},
Brendan F. Griffen\altaffilmark{1},
Phillip Zukin\altaffilmark{2,3},
Alexander P. Ji\altaffilmark{1},
Mark Vogelsberger\altaffilmark{1},
Lars E. Hernquist\altaffilmark{4},
Anna Frebel\altaffilmark{1}}

\altaffiltext{1}{Kavli Institute for Astrophysics and Space Research
  and Department of Physics, Massachusetts Institute of Technology, 77
  Massachusetts Avenue, Cambridge, MA 02139, USA; gdooley@mit.edu}
\altaffiltext{2}{Astronomy Department, University of California, Hearst Field Annex, Berkeley, CA 94720}
\altaffiltext{3}{Academia Sinica Institute of Astronomy and Astrophysics, P.O. Box 23-141, Taipei, 10617, Taiwan}
\altaffiltext{4}{Department of Astronomy, Harvard University, 60 Garden St., Cambridge, MA 02138, USA}

\begin{abstract}
We investigate how different cosmological parameters, such as those
delivered by the \textit{WMAP} and \textit{Planck} missions, affect the
nature and evolution of dark matter halo substructure. We use a 
series of flat $\Lambda$ cold dark matter
($\Lambda$CDM) cosmological $N$-body simulations of structure
formation, each with a different power spectrum but the same initial
white noise field. Our fiducial simulation is based on parameters
from the \textit{WMAP} 7th year cosmology. We then systematically
vary the spectral index, $n_s$, matter density, $\Omega_M$, and
normalization of the power spectrum, $\sigma_8$, for 7 unique simulations. 
Across these, we study variations in the subhalo mass function, mass
fraction, maximum circular velocity function, spatial distribution,
concentration, formation times, accretion times, and peak mass. We
eliminate dependence of subhalo properties on host halo mass and
average over many hosts to reduce variance. While the ``same''
subhalos from identical initial overdensity peaks in higher $\sigma_8,
n_s$, and $\Omega_m$ simulations accrete earlier and end up less
massive and closer to the halo center at $z=0$, the process of
continuous subhalo accretion and destruction leads to a steady state
distribution of these properties across all subhalos in a given host.
This steady state mechanism eliminates cosmological dependence on all
properties listed above except subhalo concentration and $V_{max}$,
which remain greater for higher $\sigma_8, n_s$ and $\Omega_m$
simulations, and subhalo formation time, which remains earlier. We also
find that the numerical technique for computing scale radius and the
halo finder used can significantly affect the concentration-mass
relationship computed for a simulation.

\end{abstract}


\section{Introduction}
The cold dark matter ($\Lambda$CDM) model of our universe has been well constrained to be flat, dark energy dominated, and filled predominantly with cold, collisionless dark matter \citep{Bennett12, Planck13}. It is partly parametrized by four quantities: the matter fraction of the universe at present day, $\Omega_m$, the primordial power spectrum scalar spectral index $n_s$, the Hubble constant at present day $H_0 = 100 h \ \mathrm{km}\ \mathrm{s}^{-1} \ \mathrm{Mpc}^{-1}$ , and the amplitude of the linear power spectrum at the scale of 8 $h^{-1}$ Mpc, $\sigma_8$. In a flat universe the dark energy content $\Omega_\Lambda$, is constrained by $\Omega_\Lambda + \Omega_M = 1$. While adherence to $\Lambda$CDM has not changed, the best estimates of these parameters have varied between recent \textit{Planck} and \textit{WMAP} measurements from \textit{WMAP} values of $\Omega_m = 0.27$, $n_s = 0.96$, $\sigma_8 = 0.80$, and $h = 0.71$ to \textit{Planck} values of $\Omega_m = 0.32$, $n_s = 0.96$, $\sigma_8 = 0.83$, and $h = 0.67$ \citep{Spergel03,Spergel07,Dunkley09, Larson11, Bennett12, Planck13}.

For over a decade now, numerical simulations adopting the $\Lambda$CDM paradigm have shown that large dark matter halos contain substructures or \textit{subhalos} which survive to the present day (e.g., \citealt{Moore99, Tormen98, Diemand07, Springel08, Giocoli10}). Various studies which attempt to connect the properties of dark matter halos to present day observables have yielded some conflicting results. There remain two key problems: the so-called ``missing satellite problem'' where there is a dearth of observed low $V_{\max}$ subhalos when compared to simulations (\citealt{Moore99}), and the ``too big to fail'' problem where there is a lack of dark ($L_V < 10^5 L_{\Sun}$), dense, high $V_{\max}$ subhalos when compared to simulations \citep{Boylan-Kolchin11}. While these issues depend critically on the influence of baryonic and radiative processes \citep{Brooks13}, as well as the possible warm \citep{Lovell13} or self-interacting nature of dark matter \citep{Vogelsberger12, Zavala13, Vogelsberger13}, a more complete understanding of the characteristics of the dominant dark matter structures is also required, including the effect of changes in cosmological parameters on halo properties. 

Recently, \cite{Angulo10} developed a technique, whose accuracy was tested by \cite{Ruiz2011}, to transform large simulation results into those of a slightly modified cosmology by adjusting length, mass, and velocity units as well as changing time time scale and amplitudes of large scale fluctuations to successfully reproduce the mass power spectrum of a given target cosmology to better than 0.5 per cent on large scales (k $<$ 0.1 $h^{-1}$ Mpc). \cite{Guo13} recently used this technique to compare \textit{WMAP}1 and \textit{WMAP}7 cosmologies on the \textit{Millennium} and \textit{Millenium-II} simulations (\citealt{Springel05}, \citealt{Boylan-Kolchin09}). Assuming a halo mass-luminosity relationship, they determined that the differences were not significant enough to be found observationally. \cite{Wang08} conducted a similar study between \textit{WMAP}1 and \textit{WMAP}3 cosmologies, but used two distinct simulations instead of transforming one into another cosmology. While the cosmologies produced minimal observable differences at low redshifts, they had to use significantly different star formation and feedback efficiencies in their models to match results to observational data. Furthermore, they concluded observational differences should be noticeable at high redshifts ($z > 3$). A number of studies have also been made of the concentration, spin and shape of dark matter halos across various WMAP cosmologies (\citealt{Duffy08}, \citealt{Maccio08}) and within the larger hierarchical framework (e.g., \citealt{Zhao09}). These studies all focused on host halos and have largely ignored subhalo properties across cosmologies. \cite{Guo13} did examine subhalos, but only their global abundances, not as a function of their host halos.
 
The dependence of host halo abundance on cosmological parameters is well understood within the Press-Schechter \citep{Press74} and improved Extended Press-Schechter (see \cite{Sheth02} for example) formalism of halo mass functions. Subhalo mass functions (SHMFs), however, are not as well predicted analytically and have not been studied extensively with respect to changing cosmologies. \cite{Zentner03} did partially investigate the effect of input cosmology on substructure using their merger history, destruction rate and survival probability as a function of spectral index, $n_s$, as well as a running spectral index model. They did not, however, investigate how these properties depend on other cosmological parameters (e.g., $\sigma_8$ or $\Omega_m$).

Several studies have investigated more general subhalo properties. \cite{Oguri04} developed an analytic model for the subhalo mass function based on the Extended Press-Schechter formalism and took account of the effects of tidal disruption and dynamical friction to estimate that the subhalo mass function is virtually independent of host halo mass. But since they used the host mass at the present day to calculate the impact of dynamical friction they inaccurately predicted the SHMF. In turn, \cite{Bosch05} examined the SHMF, mass fraction, and accretion history and found that the SHMF may not be universal, arguing that the slope and normalization depend on the ratio of the parent mass to that of the characteristic non-linear mass, M$^*$. M$^*$ indicates the typical mass scale of halos that are collapsing as a function of redshift and is defined by $\sigma(M^*,z) = \delta_c(z)$ where $\sigma(M,z)$ is the rms density fluctuation for a spherical volume of mass $M$ at redshift z, and $\delta_c(z) = 0.15(12\pi)^{2/3}[\Omega_m(z)]^{0.0055} \approx 1.68$ is the critical threshold for spherical collapse \citep{Navarro97}. The value of M$^*$ depends on cosmology and is larger for cosmologies where objects collapse sooner, i.e., higher $\sigma_8$, $n_s$ and $\Omega_m$. \cite{Bosch05} along with \cite{Giocoli10} and \cite{Gao04} however, further found that the SMHF per unit host halo mass at z = 0 is universal.

Whilst these studies have determined many of the fundamental properties of substructures, they ultimately do not systematically investigate the effect a varied cosmology has on their properties. Those that attempt to, only focus on the variance of their properties using \textit{one} cosmological parameter (e.g., \citealt{Zentner03}). In this work, we vary three key cosmological parameters ($\sigma_8$, $n_s$, $\Omega_m$) systematically and quantify the effect it has on the substructure population using simulations. In this manner, we help quantify what effect these variations have on the subhalo mass function, mass fraction, maximum circular velocity function, spatial distribution, concentration, formation times, accretion times, and peak mass in a self-consistent manner.

This paper is organized as follows. In Section \ref{sec:methods}, the simulations and halo finders used in this work are presented. Section \ref{sec:HostHalo} presents the known major effects of cosmology on host halos. In Sections \ref{sec:averaged} and \ref{sec:mergertree}, we discuss the averaged statistical properties of substructure as a function of cosmology. In Section \ref{sec:matched} we discuss the differences in substructure that is directly matched between each of our cosmological simulations. A summary of the effects of cosmology on subhalo properties is given in Section~\ref{sec:summary} and conclusions are given in Section \ref{sec:conclusion}.
\section{Numerical Methods}
\label{sec:methods}
\subsection{Simulations}
For our fiducial simulation we adopt an approximately \textit{WMAP}7 cosmology characterized by the present-day matter density parameter: $\Omega_m$ = 0.27; a cosmological constant contribution, $\Omega_\Lambda$ = 0.73; and Hubble parameter: $h$ = 0.7 ($H_0$ = 100 $h$ km s$^{-1}$ Mpc$^{-1}$). The mass perturbation spectrum has a spectral index, $n_s$ = 0.95, and is normalized by the linear rms fluctuation on 8 Mpc $h^{-1}$ radius spheres, $\sigma_8$ = 0.8. Six of our seven simulations adopt cosmologies which are identical to our fiducial run  except for individual variations in $n_s$, $\Omega_m$ and $\sigma_8$.

Cosmological initial conditions were generated at redshift z = 127 using the public code {\sc{graphic}} (\citealt{Bertschinger01}) with an Eisenstein-Hu transfer function (\citealt{Eisenstein98}). All seven simulations employed \textit{N} = 512$^3$ dark matter particles, a Plummer-equivalent comoving softening length of 1.22 $h^{-1}$ kpc, the same comoving box size, $L_\mathrm{box}$ = 25 $h^{-1}$ Mpc, and were evolved using an unreleased version of {\sc{gadget3}} (\citealt{Springel05}). Our particle masses across the seven runs range from 6.14 to 11.3 $\times$ 10$^6$ $h^{-1}$ $M_\sun$, and all simulations were written out at 64 snapshots. Our entire suite with all pertinent parameters are listed in Table ~\ref{table:simulationParameters}.

Rather than performing convergence test runs, we only investigate halos above an appropriate minimum number of particles as determined by previous studies for each halo property. We also note that our small simulation volume could lead to different absolute results than those obtained from larger simulation volumes that contain larger wavelengths of the power spectrum, particularly for halo mass functions \citep{Sirko05, Power06, Bagla09}. However, we are only concerned with the relative differences in results between cosmologies for halo substructure as a function of its host halo. In light of the conclusion by \cite{Power06} that the internal properties of CDM halos are relatively unaffected by a finite box size, it is reasonable to assume that our box's lack of larger wavelength modes does not affect any of our conclusions.

\begin{deluxetable}{ccccccc}
\tabletypesize{\footnotesize}
\tablecolumns{7} 
\tablewidth{0pt}
 \tablecaption{ Summary of the cosmological simulations
 \label{table:simulationParameters}}
 \tablehead{
 \colhead{Run} &  \colhead{$\Omega_m$} &  \colhead{$\sigma_8$} &  \colhead{$n_s$} &  \colhead{$h$} &  \colhead{$m_{p}$ ($h^{-1}$ M$_\sun$)}} 
 \startdata 
\textit{WMAP}7 & $0.27$ & $0.8$ & $0.95$ & $0.7$ & $8.72 \times 10^6$ \\
$\Omega_m = 0.35$ & {\bf 0.35} & $0.8$  & $0.95$ & $0.7$ & ${\bf 11.3 \times 10^6}$ \\
$\Omega_m = 0.19$ & {\bf 0.19} & $0.8$  & $0.95$ & $0.7$ & ${\bf 6.14 \times 10^6}$\\
$\sigma_8 = 0.9$ & $0.27$ & {\bf 0.9} & $0.95$ & $0.7$ & $8.72 \times 10^6$\\
$\sigma_8 = 0.7$ & $0.27$ & {\bf 0.7}  & $0.95$ & $0.7$  & $8.72 \times 10^6$\\
$n_s = 1.0$ & $0.27$ & $0.8$ & {\bf 1.0} & $0.7$ & $8.72 \times 10^6$\\
$n_s = 0.9$ & $0.27$ & $0.8$ & {\bf 0.9} & $0.7$ & $8.72 \times 10^6$
\enddata
 \tablecomments{$\Omega_b = 0.045$ in all runs. $\Omega_\Lambda$ + $\Omega_m$ = 1. The cosmological parameters used in the seven simulations are shown above. Each uses \textit{WMAP}7 values with one parameter (highlighted in boldface) varied at a time. }
\end{deluxetable}

\begin{figure*}
\includegraphics[width=0.98\textwidth]{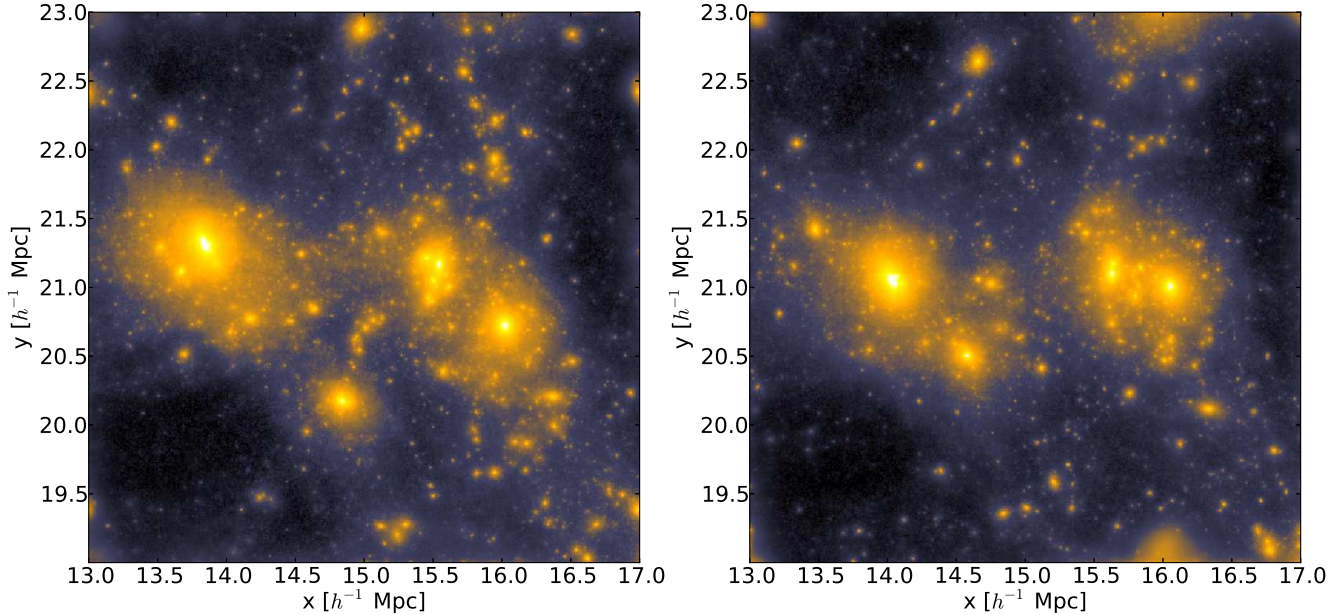}
\caption{Projected dark matter column density (log $\rho$) of corresponding halo objects in the \textit{WMAP}7 cosmology simulation with $\sigma_8 = 0.8$ at z = 0 (left panel) and the $\sigma_8 = 0.9$ simulation (right panel). The largest four halos in each visualization are matched pairs, as explained in Section \ref{sec:matching}. The cosmology with higher $\sigma_8$ results in the ``same'' halos forming earlier and merging earlier. This is visually exemplified by the pair of halos in the right side of each panel that are in the process of merging in the $\sigma_8 = 0.9$ simulation, but not in the \textit{WMAP}7, $\sigma_8 = 0.8$ simulation.}
\label{fig:matchedVisualization}
\end{figure*}

\subsection{Halo Finders}
Two different halo finders are used throughout our analysis so as to not bias our results by using a particular algorithm: {\sc{rockstar}} (\citealt{Behroozi13}) and {\sc{subfind}} (\citealt{Springel01}). As each halo finder produced the same conclusions for the relative effect of cosmology, all figures present data from {\sc{rockstar}} only. Any important systematic differences due to the halo finder are discussed in the text of the results.

$\bullet$ {\sc{rockstar}} (Robust Overdensity Calculation using K-Space Topologically Adaptive Refinement) is a phase-space halo finder which considers the position \textit{and} velocity of all particles to determine the location of halos in an $N$-body simulation. The algorithm initially selects groups of particles based on a 3D Friends-of-Friends (FOF) algorithm with a large linking length (b = 0.28). Within each FOF group, {\sc{Rockstar}} builds more FOF groups in a hierarchical fashion by adapting the linking length such that a user-specified percentage of the particles contained within a subgroup is also contained within the parent group. Once complete, these FOF groups are used to generate seed halos from the innermost level of the tree. This process is repeated until all particles at every level of the FOF group have been assigned to a halo. A further unbinding procedure is carried out to determine the final list of particles contained within each halo. For more details, see \cite{Behroozi13}.

$\bullet$ {\sc{subfind}} defines halos as locally over-dense, gravitationally bound groups of particles and begins by conducting a FOF search of the simulation volume. The local density around each particle is calculated using a smoothing Kernel over its nearest neighbors. Whenever a saddle point in a density contour which bridges two regions is reached, the smallest of the two is considered a subhalo candidate. As with the other halo finders, these candidates then undergo an unbinding procedure to produce a list of halos and subhalos. For more details, see \cite{Springel01}.

It is now well established that different halo finders have various strengths and weaknesses depending on the environment in which they are used (\citealt{Knebe11}, \citealt{Onions12}, \citealt{Knebe13}). For example, since {\sc{subfind}} is based upon an over-density criterion, its ability to identify substructure is strongly dependent on the radial position of the structure from the host (\citealt{Muldrew11}). {\sc{rockstar}} is based on a phase space algorithm and thus it does not suffer from the same problem. \cite{Onions12} found that for properties which rely on particles near the outer edge of the subhalo (e.g., total halo mass), the majority of available halo finders agree to within 10 per cent. Basic properties such as mass or the maximum circular velocity can also be reliably recovered if the subhalo contains more than 100 particles. To ensure we are not resolution limited, we only include halos and subhalos which have 300 particles or more, unless otherwise specified. 

The mass and radius of host halos used throughout this work are $r_{200}$ and $M_{200}$. $r_{200}$ is defined such that the average density of a halo within $r_{200}$ is $\bar{\rho} = \Delta_h \rho_{crit}$ where $\Delta_h = 200$, independent of cosmology as in \cite{Jenkins01}, and  $\rho_{crit}=3{H_0}^2/8\pi G$. The mass $M_{200}$ is then simply found as $M_{200} = \frac{4}{3} \pi r_{200}^3 \Delta_h \rho_{crit}$.

For subhalos, which experience tidal stripping, $m_{200}$ is a less meaningful quantity. We therefore define subhalo mass, $M_{\mathrm{sub}}$, as the total bound mass. This is computed as an extra parameter within {\sc{rockstar}}'s source code. Defining the radius of a subhalo is even more troublesome. The distance to the furthest bound particle depends on the random kinematics of a few particles, and the tidal radius is not a spherically symmetric value. Since defining a useful measure of the outer edge of a subhalo is difficult, we avoid using it. In the one analysis where it could be used, in finding subhalo concentration, we instead use a concentration definition that is independent of outer radius.

\subsection{Merger Trees}
For the merger tree component of this work (Section \ref{sec:mergertree}), we use only the merger tree generated by {\sc{rockstar consistent trees}} on the 64 snapshots per simulation. \cite{Behroozi13MT} have shown that the merger tree catalogues produced by {\sc{rockstar}} are consistent in determining halo masses, positions, and velocities when compared to merger trees constructed via different methods, e.g.,\ {\sc{bdm}} \citep{Klypin99, Klypin11} and {\sc{subfind}}. 

\subsection{Halo Matching Across Simulations}
\label{sec:matching}
Since the same white noise field was used in each simulation, particles tagged with the same {\sc{ID}} in each simulation have a low mean displacement from each other, $\sim 4$ $h^{-1}$ kpc at $z=0$. The aggregate of these similarly distributed particles are halos of similar size, position, and formation history between simulations. These ``same" halos can be matched to each other, enabling a direct assessment of how cosmological parameters affect properties of individual halos and subhalos. This analysis is carried out in Section ~\ref{sec:matched}. Figure \ref{fig:matchedVisualization} shows a visual example of matched halos in two different simulations.

Halos are matched by their particle content. In a given set of two simulations, one halo from each $\--$~a halo pair~$\--$ is compared and given a matching strength value based on the number of particles they have in common (same ID), weighted by the gravitational boundedness of each particle. For a pair of (sub)halos A and B, there are two match values, one where particles are weighted by boundedness within A, and one where particles are weighted by boundedness within B. In order for a pair to qualify as a match, (sub)halo $B$ must be the best possible match out of all (sub)halos in its simulation to (sub)halo $A$, and (sub)halo $A$ must be the best possible match to (sub)halo $B$. This dual requirement eliminates cases when a small fragment of a halo matches to a larger encompassing halo.

The particle weighting of the $i^{th}$ most bound particle (starting from $0$) in a (sub)halo is \begin{equation}
 W_i \propto \frac{n-i}{n}
\end{equation}
where $n$ is the number of particles in the halo. The total match value is given as $\sum W_i$ for all particles in common to the (sub)halo pair. In order to sum to unity for a perfect match, the normalization constant is chosen as $\frac{2n}{n+1}$. A perfect match is when all particles in (sub)halo A are found in (sub)halo B or vice-versa. Additional constraints of $\sum W_i > 0.2$ and $n > 40$ are imposed to eliminate uncertain matches. Host halos are matched first, then subhalos are matched within matched host halo pairs.

\section{Results}
This section presents results on the differences and lack of differences induced by varying cosmology on halo and subhalo properties. Effects on host halos are presented in Section~\ref{sec:HostHalo}. Section~\ref{sec:averaged} shows results for subhalo characteristics at $z=0$, Section~\ref{sec:mergertree} presents subhalo characteristics from merger tree analysis, and Section \ref{sec:matched} compares subhalo characteristics for matched subhalos.

\label{sec:results}
\subsection{Averaged Properties of Host Halos}
\label{sec:HostHalo}
While the focus of this paper is on how cosmological parameters affect substructure, we first verify and summarize for reference how cosmological parameters affect the host halo mass function and concentration.

\subsubsection{Mass Function}
\begin{figure}
\includegraphics[width=0.48\textwidth]{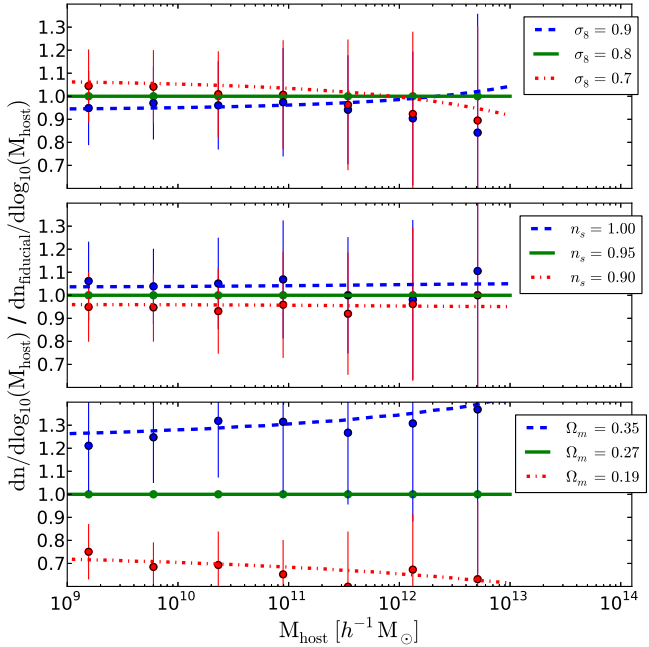}
\caption{Ratio of host halo mass functions to our fiducial simulation for different cosmological parameters. Dots correspond to data points computed from {\sc{rockstar}} halo finder data. Lines correspond to the Sheth \& Tormen analytical prediction. Error bars are computed using a jackknife method. As anticipated, in cosmologies where structure forms earlier there are more halos at $z=0$}
\label{fig:MassFunc}
\end{figure}

In accordance with \cite{Jenkins01}, we compute $\mathrm{d}n/\mathrm{d}\log M$ where $\mathrm{d}n(M)$ is the number of host halos on an infinitesimal mass interval centered at M, and $\mathrm{d}\log M$ is the logarithm of the width of the mass interval. This function shows a characteristic monotonic trend of higher abundances of low mass objects. Rather than presenting the mass function directly, we compute the ratio of the mass function at redshift $z = 0$ in each cosmology to our fiducial simulation mass function in order to accentuate any differences. This is shown in Figure~\ref{fig:MassFunc}. We also use the Extended Press-Schechter formalism developed by \cite{Sheth02} with the Eisenstein-Hu transfer function \citep{Eisenstein98} to compute analytic estimates of these mass function ratios. We compute error bars using a jackknife method \citep{Tinker08}. We create five sub-volumes, each with $\frac{4}{5}$ of the original volume, by removing a different fifth of the original volume for each sub-volume. Error bars are then the $1\: \sigma$ standard deviation of the mass function ratio as computed in each sub-volume. Our data agree with analytic estimates except on the high mass end above $10^{12}$ $h^{-1} \mathrm{M_\Sun}$, where there are fewer halos and Poisson noise becomes important.

In the top panel of Figure~\ref{fig:MassFunc}, we show that higher $\sigma_8$, which means higher amplitude for primordial density fluctuations, produces more halos above the characteristic mass scale of collapse at $z=0$, $M^* \approx 10^{13}$ $h^{-1} \mathrm{M_\Sun}$, and fewer halos below. Intuitively, a higher initial density amplitude causes more mass to end up in large halos after hierarchical merging, leaving less mass available for low mass halos. This agrees with a study by \cite{Guo13}. The reverse is true for lower $\sigma_8$.

In the middle panel, we plot the scalar spectral index, $n_s$, which controls the slope of the initial power spectrum, $P_0(k) \propto k^{n_s}$. Larger values of $n_s$ correspond to more initial power on small scales (and less on large scales), and thus more collapsed objects early on. The earlier collapsing small objects then hierarchically merge into larger objects. As evidenced by Figure~\ref{fig:MassFunc} larger $n_s$ continues to correspond to more collapsed objects by $z=0$ for the mass range considered.

We also show the effect of more exaggerated variations in $\Omega_m$. More matter content in the universe translates into objects of higher mass, and more objects, as demonstrated in the bottom panel.

Since $z = 0$ subhalos form independently of their hosts at high redshifts, these qualitative mass function differences apply just as well to subhalos before accretion. Any changes in mass and abundance are due to the dynamics of subhalo-host interactions, including tidal stripping, dynamical friction, and collisions with other subhalos.

\subsubsection{Host Concentration}
\label{sec:hostconcentration}
The density profile of dark matter halos are well approximated by the spherically symmetric Navarro, Frenk, \& White (NFW) Profile \citep{Navarro96}. The profile is defined by
\begin{equation}
\rho(r) = \rho_{crit} \frac{\delta_c}{(r/r_s)(1+r/r_s)^2},
\end{equation}
where the dimensionless $\delta_c$ is the characteristic overdensity, $r_s$ is the scale radius, and $\rho_{crit}$ is the critical density of the universe. Concentration, defined by $c_{200} \equiv r_{200}/r_s$, gives a measure of how centrally concentrated the particles in the halo are. Concentration could also be defined in terms of the spherically symmetric Einasto density profile, given as
\begin{equation}
\rho(r) = \rho(r_{-2}) \exp \left[-\frac{2}{\alpha} \left( \left(\frac{r}{r_{-2}} \right)^{\alpha}-1 \right)  \right],
\end{equation}
where $r_{-2}$ is the radius at which the logarithmic slope of the profile is $-2$, the shape parameter $\alpha$ is an extra free parameter to fit to, and $\rho(r_{-2})$ is the density at $r_{-2}$ \citep{Einasto65}. In this case the concentration would be $c_{200} \equiv r_{200}/r_{-2}$. Furthermore, many studies use outer radii other than $r_{200}$. Since each definition quantifies the same qualitative idea, we consider only the first definition in terms of the NFW profile. We use the default $r_{200}$ as determined by the halo finder, but compute $r_s$ directly from the particle data of the halo. To facilitate a direct comparison to past studies, we find $r_s$ by fitting an NFW profile to the density profile of each halo, determined by binning particles in 16 bins equally spaced in log space from $\mathrm{log}_{10}(\frac{r}{r_{200}}) = -1.25$ to $0$, as in \cite{Duffy08}.

Host halo concentrations have been studied extensively (e.g., \citealt{Zhao09}). It is now well established that halos which collapse earlier have higher concentrations (\citealt{Neto07}, \citealt{Duffy08}, \citealt{Maccio08}, \citealt{Giocoli10}). This is simply because the universe was of higher density at early times. One would then expect cosmologies with a higher $M^*$ to have higher concentrations. Similarly, halos with a higher mass should have smaller concentrations since they formed later. In particular, for increasing values of $\sigma_8$ we expect the concentration to increase since it reflects the background density of the universe at the time when the halo formed.

To test this, we consider all ``relaxed'' halos whose hosts satisfy the following criteria similar to \cite{Neto07}:
\begin{itemize}
\item The host must contain at least 600 particles.
\item The substructure mass fraction of a given host must be, $f_{s} < 0.1$.
\item The offset parameter or center of mass displacement must be s $<$ 0.07 where $s = |r_c - r_{cm}|/r_{200}$. $r_c$ and $r_{com}$ are the center of halo peak density and mass respectively.
\item The virial ratio, $2T/|U|$, must be less than a pre-set value of $1.35$, where $T$ is the total kinetic energy and $U$ is the gravitational potential self-energy of the host halo.
\end{itemize}

\begin{figure}[!h]
\includegraphics[width=0.48\textwidth]{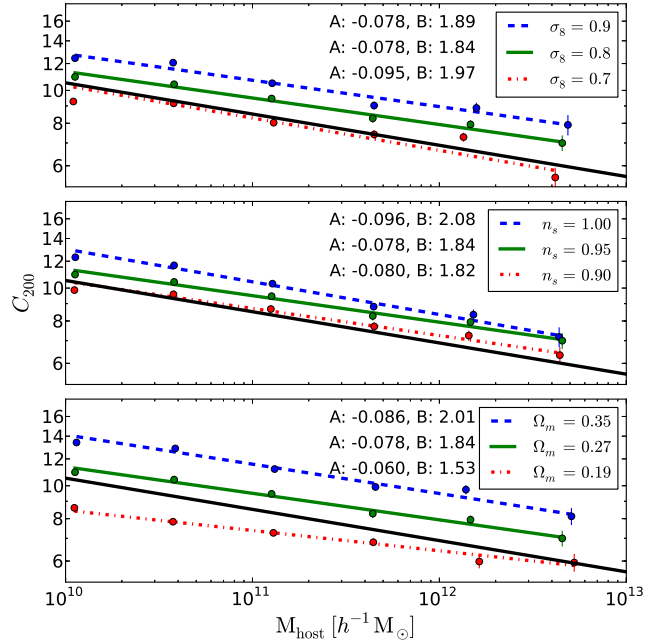}
\caption{The averaged host halo concentration as a function of host halo mass and cosmology. Best fit lines and best fit parameters as described in Eq. \ref{equationCM}, and error of the mean bars are shown. The $c\--M$ relation derived by \cite{Duffy08} (solid black line) is consistent in its slope to our findings. Its overall magnitude, however, is lower primarily due to its computation using {\sc{subfind}} and our data using {\sc{rockstar}}. Cosmologies with higher $M^*$ lead to more concentrated halos, but do not affect the slope of the $c\--M$ relation.}
\label{fig:hostchostM}
\end{figure}

This combination of mass and kinematic information encoded in the substructure mass fraction, offset parameter, and virial ratio ensures that the hosts in our relaxed sample are well fitted by NFW profiles ~\citep{Navarro96}. Unrelaxed halos are more difficult to accurately fit a scale radius to. Figure~\ref{fig:hostchostM} shows the concentration-mass ($c\--M$) relation for each of our simulations. Using the mean and rms deviation of $\mathrm{log}_{10}(c_{200})$, we fit the binned $c\--M$ relation using,

\begin{equation}
\mathrm{log}_{10}(c_{200}) = A \ \mathrm{log}_{10}{\left( M_{vir} \right)} + B,
\label{equationCM}
\end{equation}
where log$_{10}(c_{200})$ and $M_{vir}$ are the mean values in each bin. For the error of the mean in each bin, $\sigma_c$, we use the rms deviation of log$_{10}(c_{200})$ divided by the square root of the number of objects in each bin. For each fit, we define
\begin{equation}
\chi^2 = \sum_{j=1}^{N_{b}}{\left(\frac{\mathrm{log}_{10}(c_{200}) - \mathrm{log}_{10}(c_{200fit_j})}{\sigma_{\bar{c}_j}}\right)^2},
\end{equation}
where $N_{b}$ is the number of bins over which the fit is performed and $c_{200fit_j}$ is obtained from the best fit of Equation (\ref{equationCM}), and find the values of $A$ and $B$ such that the reduced chi-squared, $\bar{\chi}^2$ ($\chi^2$ divided by the number of degrees of freedom), is minimized. All error bars and best fit functions in future figures are found using this same method.

Our data and best fit functions for the relaxed sample are shown in Figure~\ref{fig:hostchostM}. We also add for comparison the relation for $c_{200}$ derived by \cite{Duffy08} who used \textit{WMAP}5 data ($\sigma_8 = 0.796$, $n_s = 0.963$, $\Omega_m = 0.258$). The data confirms that higher $M^*$ cosmologies and smaller host halos do in fact have higher concentrations. The slopes of each $c\--M$ relation range from $-0.060$ to $-0.097$ with a typical value of $-0.083$, but do not vary systematically with cosmological parameters. Furthermore, the slope is sensitive to within $\pm 0.008$ to how the host halos are binned and the halo mass range considered. Therefore we can only definitively confirm how the amplitude of the $c\--M$ relation is affected by cosmological parameters, and not the slope. The slopes obtained do agree, within error, with the slope of $-0.092$ obtained by \cite{Duffy08} for relaxed halos.

We further note that the slope and magnitude of the $c\--M$ relation is very sensitive to the method for finding $r_s$ and the halo finder. By default, {\sc{rockstar}} computes $r_s$ by fitting an NFW profile to a density profile found by dividing particles into 50 radial bins such that each bin contains the same mass \citep{Behroozi13}. This results in a much steeper slope for all cosmologies, with a typical slope of $-0.15$. Furthermore, the concentrations computed using {\sc{rockstar}}'s $r_{200}$ and particle assignments are systematically higher than those found using {\sc{subfind}}'s parameters. Using {\sc{subfind}}, our fiducial simulation's $c\--M$ relation is very similar to the \cite{Duffy08} relationship in magnitude and slope, which was also computed using {\sc{subfind}}. The $c\--M$ relationships found using {\sc{rockstar}}, however, are on average 12\% greater in overall magnitude. Both halo finders do, however, agree on the relative differences between cosmologies. The large dependencies of concentration on its method of computation are examined in greater detail by \cite{Meneghetti13}.

\subsection{Averaged Properties of Subhalos}
\label{sec:averaged}
Subhalo abundances and properties depend dramatically on the size of their host halo. Any useful comparison between simulated and observed subhalo distributions thus requires comparing distributions within hosts of the same size. In this section, we explore the effects of cosmology on averaged subhalo properties as a function of their host halo mass.

\subsubsection{Subhalo Mass Function}
Similar to the host halo mass function, the subhalo mass function, SHMF, counts subhalo abundance within a chosen host halo as a function of mass. Whereas the host halo mass function has been studied extensively by numerical simulations and agrees well with fully analytic predictions, see Figure~\ref{fig:MassFunc}, only semi-analytic models exist for the SHMF \citep{Oguri04, Bosch05, Springel08, Zentner03, Gao04}. This is due to the more complex effects of collisions, dynamical friction, and tidal stripping changing subhalo masses over time.

\begin{figure}[!h]
\includegraphics[width=0.48\textwidth]{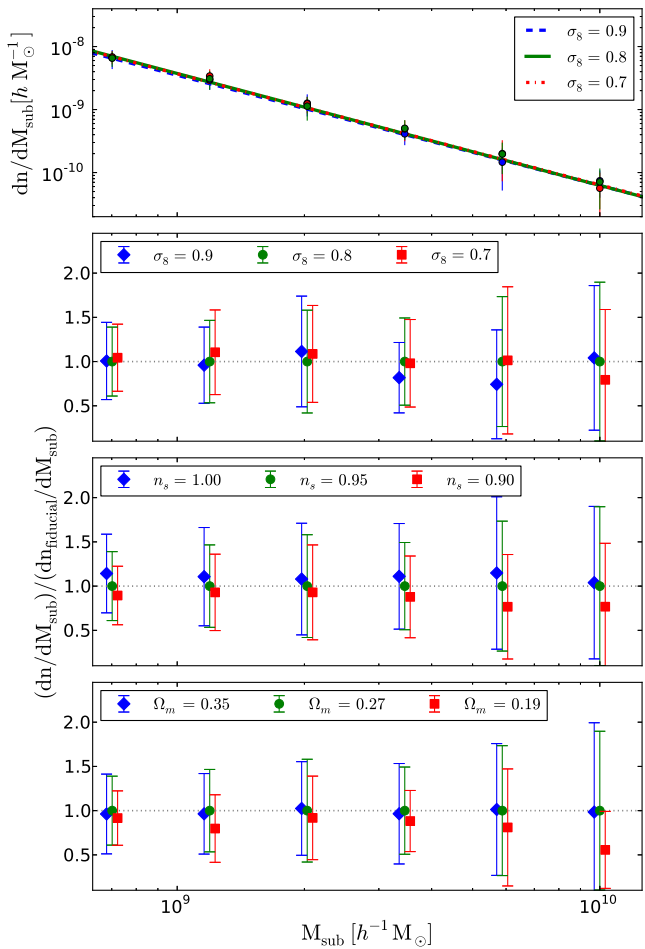}
\caption{\emph{Top Panel:} Subhalo mass function normalized to host halo mass and scaled to a Milky Way sized host of $0.84 \times 10^{12}$ $h^{-1}\mathrm{M_\Sun}$ ($1.2 \times 10^{12}$ $\mathrm{M_\Sun}$), averaged over all host halos above $10^{12.5}$ $h^{-1}\mathrm{M_\Sun}$ for cosmologies with different $\sigma_8$. Resolution constraints on low masses and large Poisson errors for high masses constrain the subhalo mass range considered. Best fit lines and $1\: \sigma$ halo-to-halo variation bars, not error bars, are shown. \emph{Bottom Panels:} Ratio of subhalo mass function to the fiducial \textit{WMAP}7 subhalo mass function. Data points for the non \textit{WMAP}7 cosmologies are shifted slightly left and right of their true values to help distinguish them. The effects of varying $\sigma_8$ and $\Omega_m$ are consistent with no change in the SHMF. A small trend of higher abundance with higher $n_s$ exists but is also consistent with no change in the SHMF within error. Variation on a halo-to-halo basis dominates any effects of cosmology on the SHMF as seen in the $1\: \sigma$ variation bars.}
\label{fig:RS_SHMF}
\end{figure}

We investigate the SHMF, $\mathrm{d}n/dM_{sub}$ vs $M_{sub}$, where $\mathrm{d}n$ is the number of subhalos in a mass interval $\mathrm{d}M_{sub}$. Within each particular cosmology, we find that the SHMF per unit host halo mass (i.e., dividing the differential abundance by the mass of the host halo), yields a universal function. This confirms the same result found in \cite{Gao04} and \cite{Bosch05}. We use this trait to average the normalized subhalo mass functions for all hosts halos with $M_{host} > 10^{12.5}$ $h^{-1}\mathrm{M_\Sun}$ to account for halo-to-halo variance of the SHMF. Lowering the $M_{host}$ mass threshold simply increases variance, due to resolution effects, without changing the value of the average mass function. The top panel of Figure~\ref{fig:RS_SHMF} shows the SHMF per unit host halo mass scaled for a Milky Way sized host of $M_{host} = 0.84 \times 10^{12}$ $h^{-1}\mathrm{M_\Sun}$ ($1.2 \times 10^{12}$ $\mathrm{M_\Sun}$). We also indicate the magnitude of $1\: \sigma$ halo-to-halo variation with bars on each point to compare changes due to cosmology with inherent variance between halos. This strategy of handling variance by finding characterizations that are independent of host halo mass, then averaging the characterizations over all host halos, is used throughout the remainder of Section~\ref{sec:averaged}.

\begin{figure}
\includegraphics[width=0.48\textwidth]{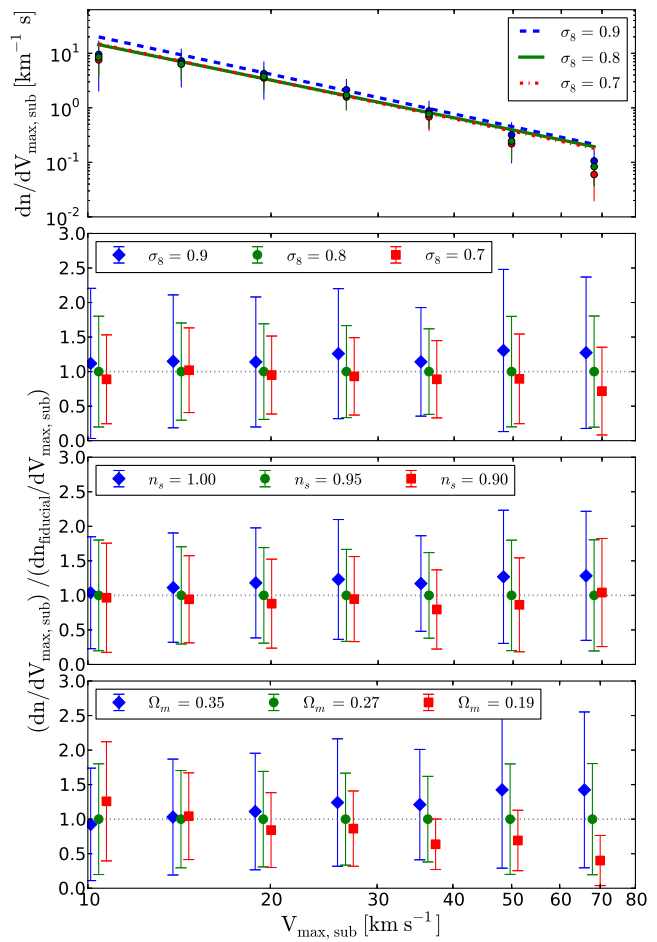}
\caption{\emph{Top Panel:} Subhalo $V_{max}$ function normalized to host halo $V_{max}$ and scaled to a Milky Way sized host of $220\ \mathrm{km\ s^{-1}}$, averaged over all host halos above $10^{12.5}$ $h^{-1}\mathrm{M_\Sun}$ for cosmologies with different $\sigma_8$. Best fit lines and $1\: \sigma$ halo-to-halo variation bars are shown. \emph{Bottom Panels:} Ratio of subhalo $V_{max}$ function to the fiducial \textit{WMAP}7 subhalo $V_{max}$ function. Data points for the non \textit{WMAP}7 cosmologies are shifted slightly left and right of their true values to help distinguish them. Higher values of each of the cosmological parameter (higher $M^*$) result in higher abundances of $V_{max}$ subhalos over the mass range considered. The cosmologies chosen correspond to $\sim 18\%$ greater abundance for the higher $M^*$ cosmologies and $\sim 10\%$ less abundance for the lower $M^*$ cosmologies. Variation on a halo-to-halo basis dominates any effects of cosmology on the $V_{max}$ function as seen in the $1\: \sigma$ variation bars.}
\label{fig:RS_SHVF}
\end{figure}

Varying the parameters $\sigma_8$ and $\Omega_m$ has no noticeable effect on the subhalo abundance at $z=0$. Our simulations with higher values of $n_s$ have a greater abundance of subhalos in each bin. However, the error on this overall trend is equal to the magnitude of the trend so it may have arisen due to chance. Additionally, variations from halo-to-halo in the SHMF exceed any possible small cosmological effect. The best fit to the data produced from {\sc{rockstar}} data is given by
\begin{equation}
\frac{\mathrm{d}n}{\mathrm{d}M_{sub}} = K \times \left(\frac{M_{sub}}{h^{-1} \mathrm{M_\Sun}} \right)^{-\alpha}\frac{M_{host}}{h^{-1}\mathrm{M_\odot}}
\label{eq:SHMF}
\end{equation}
Where $K = 4.5 \pm 0.3 \times 10^{-5} h\ \mathrm{M_\Sun^{-1}}$ and $\alpha = 1.78 \pm 0.04$. The same results and fit values were reproduced within error using {\sc{subfind}}, the halo finder used in most previous studies. The power law exponent of $-1.78$ falls between the values of $-1.73$ reported by \cite{Helmi02} and $-1.9$ reported by \cite{Springel08}, \cite{Bosch05} and \cite{Gao04}. Individual halos in this study have SHMFs with $\alpha$ ranging from $1.65$ to $1.95$, encompassing the range of values reported in the literature.

Following from an unchanging SHMF per host halo mass, the subhalo mass fraction is also not changed by variations in cosmological parameters. This was directly confirmed by simulation data.

\subsubsection{Maximum Circular Velocity}
While mass may be the most intuitive description of a subhalo's size, the maximum circular velocity, $V_{max}$, is a related measure that is easier to ascertain observationally and more robust to measure from simulations. It is defined as the maximum velocity of an orbiting body in the potential of a halo:

\begin{equation}
V_{max} = \mathrm{max} \left(\frac{GM(<r)}{r}\right)^{\frac{1}{2}}.
\end{equation}

Unlike mass, $V_{max}$ is not sensitive to the poorly defined boundary between the halo and the background \citep{Kravtsov10}. Additionally, it avoids the arbitrariness that plagues any definition of mass. While closely related, $V_{max}$ does not just trace mass, but also has a weak dependence on concentration: higher concentration leads to higher $V_{max}$ \citep{Bullock01}. Thus, if there is no cosmology dependence on the SHMF, there should in principle still be a weak dependence on $V_{max}$ due to the dependence of subhalo concentration on cosmology (see Figure~\ref{fig:hostmvirsubc}). For these reasons, we repeat the subhalo abundance analysis done for subhalo mass with subhalo $V_{max}$ instead. The subhalo $V_{max}$ function is computed for each host halo, normalized to the host's $V_{max}$, averaged over all host halos with $M_{host} > 10^{12.5}$  $h^{-1}\mathrm{M_\Sun}$, then scaled to a host with $V_{max} = 220\ \mathrm{km \ s^{-1}}$. The resulting function and the ratio of it to our fiducial simulation's function for each cosmology are shown in Figure~\ref{fig:RS_SHVF}. We indeed find that this $V_{max}$ function depends weakly on the cosmological parameters considered. The greater value of each cosmological parameter increases the abundance of subhalos as a function of $V_{max}$ by $\sim 18\%$ and the lower values of each parameter decreases abundance by $\sim 10\%$. Still, the error on this trend is substantial at $\sim 10\%$, and variation from halo to halo exceeds this effect.

Just as for the SHMF, we fit an exponential function to the data. This equation is
\begin{equation}
\frac{\mathrm{d}n}{\mathrm{d}V_{max,\ sub}} = K_v \times \left(\frac{V_{max,\ sub}}{\mathrm{km\ s^{-1}}} \right)^{-\alpha_v}\frac{V_{max,\ host}}{\mathrm{km\ s^{-1}}}
\label{eq:SHVF}
\end{equation}
and the best fit parameters to the fiducial simulation are $K_v = 12.5 \pm 0.9\ \mathrm{km^{-1}\ s}$ and $\alpha_v = -2.3 \pm 0.2$.
 
\begin{figure}[!h]
\includegraphics[width=0.48\textwidth]{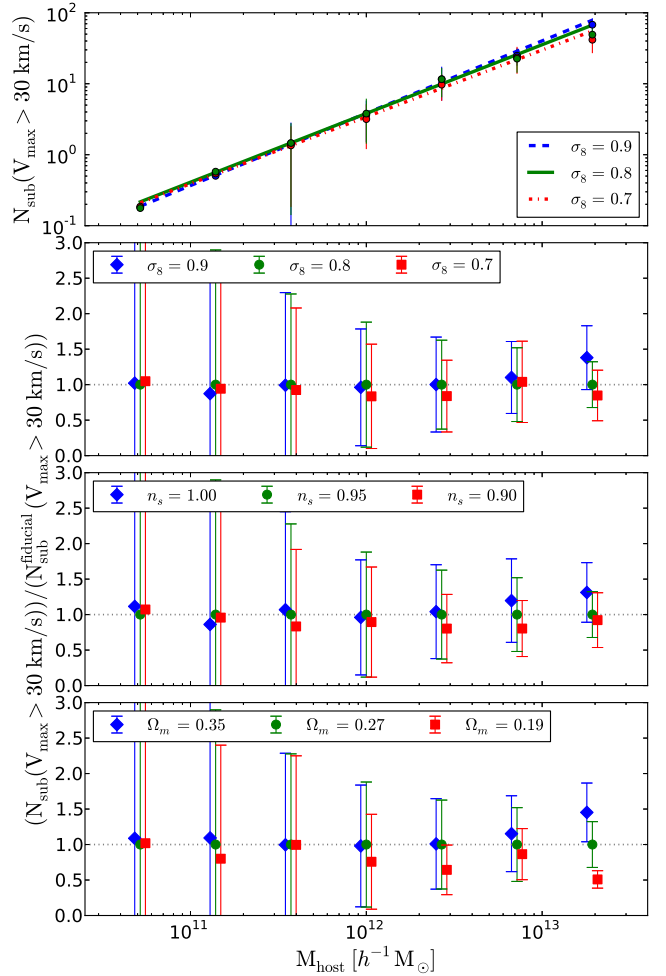}
\caption{\emph{Top Panel:} Average number of subhalos with $V_{max} > \mathrm{30 \ km/s}$ as a function of host halo mass and cosmology for variations in $\sigma_8$. The number of subhalos above this threshold is not affected by cosmology, but is substantially affected by the mass of the host halo. Halo-to-halo variation is shown in the $1\: \sigma$ variation bars. \emph{Bottom Panels:} Ratio of the number of subhalos with $V_{max} > \mathrm{30 \ km/s}$ in a given cosmology to that of the fiducial \textit{WMAP}7 cosmology. Data points for the non \textit{WMAP}7 cosmologies are shifted slightly left and right of their true values to help distinguish them. All changes in cosmology tested are consistent with having no effect on this function. Variation on a halo-to-halo basis dominates any effects of cosmology as seen in the $1\: \sigma$ variation bars.}
\label{fig:vmax}
\end{figure}

$V_{max}$ is of particular interest in relation to the Missing Satellite Problem. \cite{Wang12} discuss the disparity between subhalo abundance above $V_{max} = 30 \ \mathrm{km \ s^{-1}}$ observed in the Milky Way and the abundance predicted in simulations of a $10^{12}$ $h^{-1} \mathrm{M_{\Sun}}$ halo. The Aquarius simulations predict eight subhalos, whereas the Milky Way has only three with $V_{max}$ above $30 \ \mathrm{km \ s^{-1}}$ (the Large  Magellanic Cloud, the Small Magellanic cloud, and the Sagittarius Dwarf). They discover that this number is very sensitive to the host halo mass and very insensitive to the cosmology. With the data to test this more explicitly, we show in Figure~\ref{fig:vmax} the average number of subhalos with $V_{max} > \mathrm{30 \ km\ s^{-1}}$ as a function of host halo mass. We find that this $V_{max}$ function is consistent with being independent of the variation in cosmological parameters considered, confirming the result of \cite{Wang12}. 

\subsubsection{Spatial Distribution}
The subhalo spatial distribution in terms of its number density as a function of radius is also studied. Following \cite{Springel08}, we first investigate the number density profile, $n(r)/\!\!<\!\!n\!\!>$, where $n(r)$ is the local number density of subhalos, and $<\!\!n\!\!>$ is the average subhalo number density within the virial radius. With the radial distance normalized to $r_{200}$, we discover that this distribution takes on a characteristic form for all host halos, regardless of mass, for $r/r_{200} > 0.4$. We tested for this independence of mass by computing the average distribution over several host halo mass intervals in the range $10^{10}$ $h^{-1}\mathrm{M_{\Sun}}< M_{host} < 10^{13.6}$ $h^{-1}\mathrm{M_{\Sun}}$. No mass dependent difference was found in the profiles above $r/r_{200} > 0.4$. Below this threshold, corresponding to only $6.4\%$ of the halo volume, there is a clear mass dependent trend with smaller host halos having higher number densities than larger host halos. However, this region is also one where halo finders have difficulty identifying subhalos, so it is unclear whether the effect is real, or a halo finding artifact. We resolve to exclude this range from the cosmology comparison analysis. 

\begin{figure}[!h]
\includegraphics[width=3in]{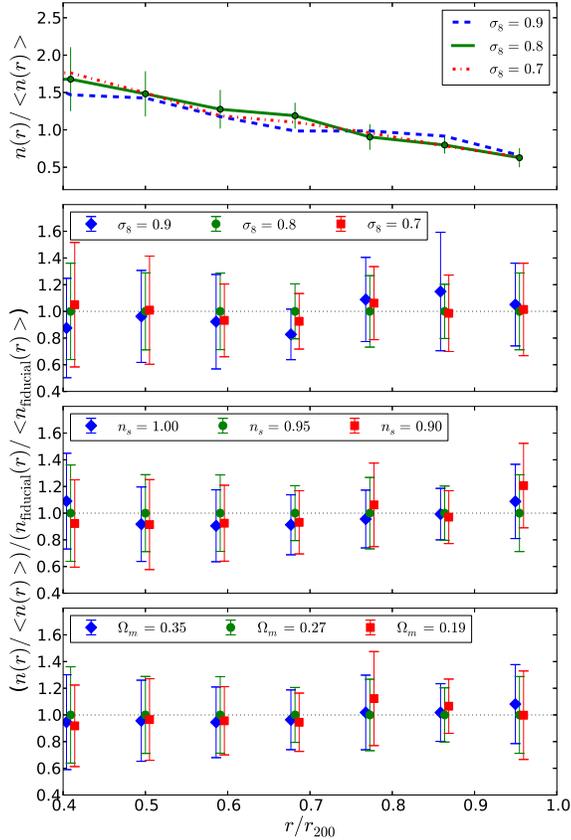}
\caption{\emph{Top Panel:} The averaged subhalo number density as a function of $r/r_{vir}$ normalized to the mean subhalo number density for cosmologies with varied $\sigma_8$. This function is found to not vary with the mass of the host halo nor the cosmological parameters considered for $r/r_{vir} > 0.4$. At smaller radii, halo finding becomes too uncertain to make any claim. Variation bars show $1\: \sigma$ variations in the density profiles of individual halos. \emph{Bottom Panels:} Ratio of the subhalo spatial distribution as in the top panel to that of the to the fiducial \textit{WMAP}7 cosmology for all cosmologies. Data points for the non \textit{WMAP}7 cosmologies are shifted slightly left and right of their true values to help distinguish them. All changes in cosmology tested are consistent with having no effect on this function. Variation on a halo-to-halo basis dominates any effects of cosmology on subhalo spatial distribution as seen in the $1\: \sigma$ variation bars.}
\label{fig:numdensprofile}
\end{figure}

Figure~\ref{fig:numdensprofile} shows the characteristic subhalo radial distribution averaged over all host halos with $M_{host} > 10^{12.5}$ $h^{-1}\mathrm{M_\Sun}$ for each cosmology. Once again, this distribution is found to be independent of the cosmological parameters studied. 

\subsubsection{Subhalo Concentration}
Using similar methodology presented in Section \ref{sec:hostconcentration}, we examine the subhalo $c\--M$ relation for subhalos within relaxed hosts at redshift $z = 0$. Due to difficulty in defining subhalo mass, we also explored the subhalo $c\--V_{max}$ relation and found all of the same results. Since both options lead to the same conclusions, we choose to present only the $c\--M$ relation in order to compare to previous literature and the host halo $c\--M$ relation. The concentration, $c$, is once again defined by the ratio $c \equiv r_{200}/r_s$. However, directly computing $r_{200}$ and $r_s$ for subhalos is troublesome. As noted in Section~\ref{sec:hostconcentration}, the $c\--M$ relation is very sensitive to how $r_s$ in computed. Whereas for host halos the method used by \cite{Duffy08} of dividing particles into 16 logarithmic bins to make a density profile, then fitting an NFW profile to find $r_s$ works consistently well, this method does not work reliably for subhalos which typically have too few particles to populate all 16 bins. The default method from {\sc{rockstar}} overcomes this issue by choosing 50 bins such that each contains an equal number of particles. However, we find that this method results in erratically varying subhalo concentrations with standard deviations of concentration per subhalo mass interval exceeding $80$.

We therefore avoid computing $r_s$ directly and instead use other parameters and properties of an NFW profile to infer $c$. \citep{Klypin11} uses $V_{max}$ and $M_{200}$ to numerically solve for $c$ assuming an NFW profile. While we did test out this method, $M_{200}$ is nonsensical for very tidally stripped subhalos where $r_{200}$ exceeds the furthest bound particle. This occurs in $\approx 5\%$ of subhalos above $10^9$ $h^{-1}\mathrm{M_{\Sun}}$. We thus finally turn to the method used by \cite{Springel08} which also assumes an NFW profile, but uses $V_{max}$ and the radius where $V_{max}$ occurs, $r_{vmax}$, to compute $c$. The concentration is found numerically from the equation
\begin{equation}
\frac{c^3}{\ln(1+c) - c/(1+c)} = .21639 \left( \frac{V_{max}}{H_0 r_{vmax}} \right)^2.
\label{eq:subhalo_c}
\end{equation}

\begin{figure}[!h]
\includegraphics[width=0.48\textwidth]{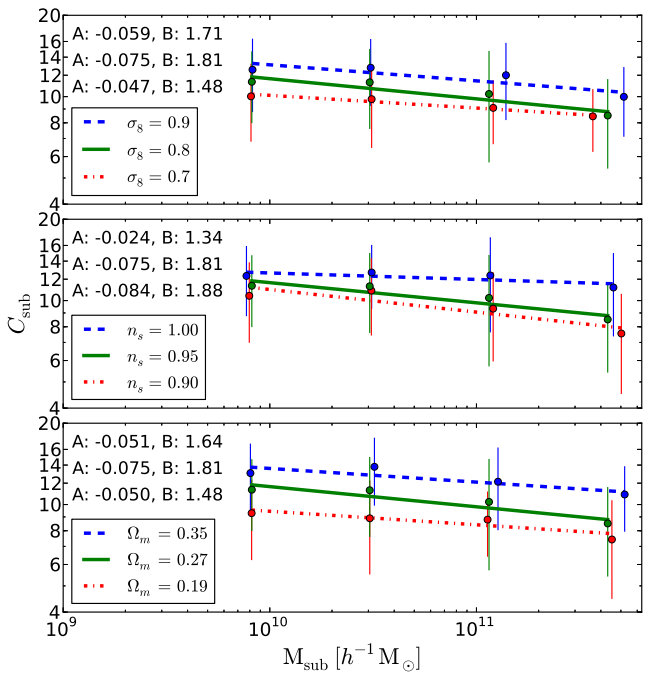}
\caption{The averaged subhalo concentration as a function of subhalo mass and cosmology. Concentration is computed using $r_{vmax}$ and $V_{max}$ according to Eq.~\ref{eq:subhalo_c} which assumes an NFW profile. The differences in concentration due to cosmology seen in the host halos remain in the subhalos. Cosmologies with higher $M^*$ lead to subhalos with higher concentration. The same trend holds for the $c\--V_{max}$ relationship. The magnitude of the subhalo concentrations cannot be reliably compared to the host halo concentrations of Figure~\ref{fig:hostchostM} since they were computed differently. Variation bars show large $1\: \sigma$ variations in the concentration of individual subhalos. Best fit lines and best fit parameters as described in Eq. \ref{equationCM} are also shown. Difficulty in computing concentration reliably below 500 particles per subhalo sets the low mass cut-off in the plot.}

\label{fig:csubmsub}
\end{figure}

The $c\--M$ relation from this method for all subhalos with greater than 500 particles is shown in Figure~\ref{fig:csubmsub}. As a function of cosmology, we discover that the differences in concentration demonstrated in Figure~\ref{fig:hostchostM} still last in the subhalo population. Cosmologies with higher $M^*$ (higher $\sigma_8, n_s$, and $\Omega_m$) lead to subhalos with higher concentration. We further fit the form of Eq.~\ref{equationCM} to the data and find concentrations similar to the host halos but with more shallow negative slopes. Due to a more limited sample of subhalos than host halos after the selection criteria, $\sim 900$ vs. $\sim 6000$ in the $10^{9.64} - 10^{12}$ $h^{-1}\mathrm{M_\Sun}$ interval, and density profiles that deviate further from NFW, the strength of each fit is worse and the variation in slopes between simulations is much greater for subhalos. Furthermore, using the Klypin method of computing subhalo concentration yields steeper negative slopes than the host halo $c\--M$ relation (it yields a median slope of $-0.122$, in agreement with the slope of $-0.12$ found by \cite{Klypin11}), and a $70 - 95\%$ increase in concentration relative to hosts for the mass range considered. We therefore cannot definitively claim a characteristic slope value or magnitude for subhalo concentrations and caution that results are very sensitive to the method used to compute halo concentration. The Klypin method does nonetheless agree with the relative differences between cosmologies, and we conclude that varying cosmology does have a real effect on subhalo concentrations. 

With a consistent computation of concentration for host halos and subhalos according to $Eq.~\ref{eq:subhalo_c}$, we find that the median slope of the $c\--M$ relation for hosts is $-0.06$, which is consistent with the slopes found for the subhalos. The concentration of subhalos in the same mass intervals is typically $\approx 15\%$ greater. Merger tree histories of subhalos demonstrate that their concentrations do in fact continually rise after accretion, and at a faster rate than host halos. This can be explained by a change in subhalo density profiles due to tidal stripping and tidal heating. \citet{Hayashi03} models the density profiles as a modified NFW function that changes over time as a function of the ratio of current subhalo mass to its mass at infall. The modified NFW profile and supporting fitting functions are given in \citet{Hayashi03} as equations 8, 9, and 10. Following this model, both $V_{max}$ and $r_{vmax}$ decrease as a function of mass loss, and thus time since accretion. However, since $r_{vmax}$ decreases faster than $V_{max}$, subhalo concentration as computed in Eq~\ref{eq:subhalo_c} increases monotonically. These trends are confirmed in the merger tree histories of the subhalos, but with significant variance. Since the subhalo profile is no longer NFW in this model, Eq~\ref{eq:subhalo_c} only approximates the original definition of concentration, $c = r_{200}/r_s$. We therefore also study the poorly defined $r_s$ and $r_{200}$. This similarly shows $r_s$ decreasing at a faster rate than $r_{200}$. Below the limit of 500 particles, the concentration of subhalos and host halos systematically and unrealistically decreases as a function of mass, but the relative differences between cosmologies remains.

\begin{figure}[!h]
\includegraphics[width=0.48\textwidth]{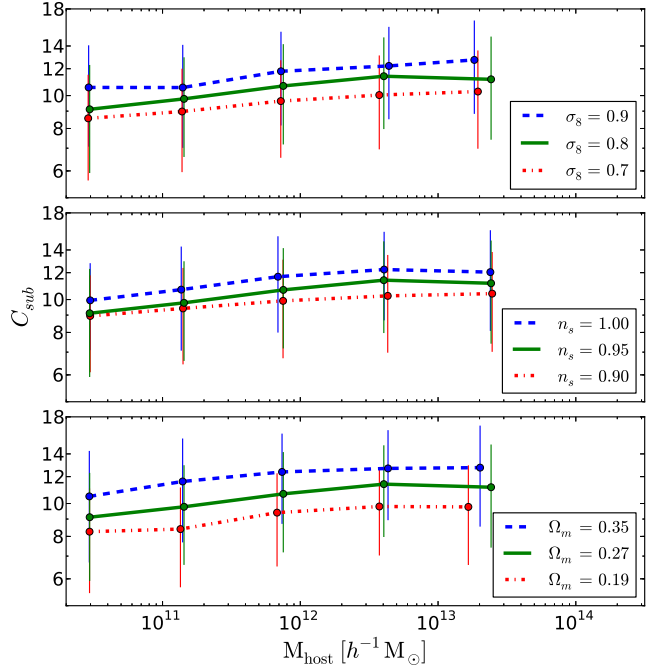}  
\caption{The averaged subhalo concentration as a function of host halo mass and cosmology for subhalos between $10^{9}$ and $10^{10}$ $h^{-1}\mathrm{M_{\Sun}}$ in relaxed hosts. The slight positive trend suggests that larger host halos cause a greater increase in subhalo concentration than smaller host halos, possibly due to stronger tidal forces. Without a theory to predict a form of this trend, no function is fit to the data. Variation bars show large $1\: \sigma$ variations in the concentration of individual subhalos.}
\label{fig:hostmvirsubc}
\end{figure}

We further investigate the subhalo concentration as a function of host mass in Figure~\ref{fig:hostmvirsubc}. We find that the average subhalo concentration is weakly dependent on the host mass, with larger hosts having more concentrated subhalos. We account for the bias of larger hosts having a different distribution of subhalos by only considering subhalos in $d \log_{10}(M_{sub}) = 0.5$ mass intervals from $10^8$ to $10^{11}$ $h^{-1}\mathrm{M_{\Sun}}$. In every case we observe the same positive slope. We also rule out the possibility that larger hosts tend to have halos which were accreted earlier through studies of the merger tree history, as presented in Sections~\ref{sec:Formation} and \ref{sec:Accretion}. Without a strong theoretical motivation for any functional form, we abstain from making a best fit. We speculate that the trend may be due to stronger tidal forces in larger hosts. This would indicate that the subhalo profile models from \citet{Hayashi03} should depend weakly on the host halo's size in addition to the fraction of remaining bound subhalo mass. A full explanation is beyond the scope of this paper.

\subsection{Merger Tree Analysis}
\label{sec:mergertree}
While the previous subsection focused on static distributions at $z=0$, we now investigate the evolution of subhalos as a function of cosmology through merger tree analysis. Halos that become subhalos go through a trajectory of forming, gaining mass, reaching a maximal mass, losing mass due to tidal stripping from a nearby halo, entering $r_{200}$ of their eventual host (accretion), and ultimately merging with their host through stripping and disruption. Subhalos detected at $z=0$ are ones that have not yet been fully torn apart by their host. The following sections explore these steps with the approach of finding average distributions for subhalos within a host. Rather than presenting results for all cosmologies, we highlight only simulations with varied $\sigma_8$, which has historically been the least well constrained parameter of the three studied in this paper.

\subsubsection{Formation Time of Subhalos}
\label{sec:Formation}
Numerous definitions exist for the formation time of dark matter halos. As discussed in \cite{Li08}, these definitions generally fall into two classes:
\begin{enumerate}
\item When a halo reaches a fraction of its final mass or $V_{max}$.\\
\item When a halo first reaches a threshold mass or  $V_{max}$.
\end{enumerate}
While both types of definitions can be applied to host halos, the first class does not apply well to subhalos. Subhalos do not monotonically grow in mass, but rather reach a peak before the effects of tidal stripping from a nearby host removes mass. Thus, the time when a subhalo reaches a fraction of its final or maximal mass intertwines its formation and post-accretion history. We therefore use the definition that a halo forms once it first surpasses a threshold mass of $M_{200} = 3 \times 10^8$ $h^{-1} \mathrm{M_\Sun}$, corresponding to $\sim 35$ particles. Changing the mass threshold shifts when halos have ``formed'', but does not influence how cosmology affects formation times. This mass was chosen to minimize the number of halos which were first detected above the threshold mass, and those which form but never reach the threshold mass. It succeeds in characterizing $\sim 55\%$ of subhalos at $z=0$. The remaining $45\%$ of subhalos typically have incomplete mass histories, so their formation time would be hard to deduce under any definition of formation time. The threshold mass definition has the useful property that the formation time of subhalos is a tracer of formation time of the oldest stars in each subhalo \citep{Li08}.

\begin{figure}[!h]
\includegraphics[width=0.48\textwidth]{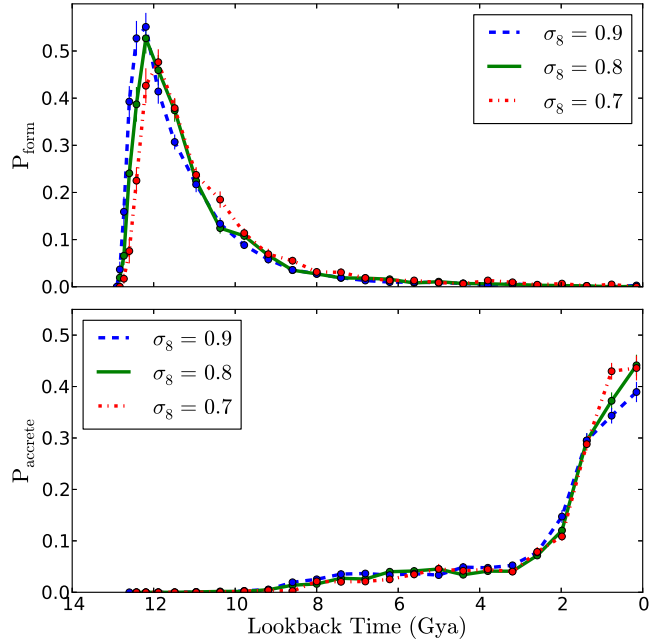}
\caption{\emph{Top Panel}: Average formation time probability density distribution of $z=0$ subhalos within a host for cosmologies with varied $\sigma_8$. Formation time defined by when halos first reach a mass of $M_{200} = 3 \times 10^8$ $h^{-1} \mathrm{M_\Sun}$. Subhalos lasting to $z=0$ typically formed earlier in higher $\sigma_8$ cosmologies by $.1$ to $.4$ billion years for a $0.1$ increment in $\sigma_8$. \emph{Bottom Panel}: Average accretion time probability density distribution of subhalos within a host for cosmologies with varied $\sigma_8$. No significant differences are found between the cosmologies. Error bars show $1\: \sigma$ uncertainty of each data point.}
\label{fig:tform_accrete}
\end{figure}

We use the merger trees to study the probability density distribution of subhalo formation times for all surviving subhalos at $z=0$ within a host. This distribution is similar for all host halos in a simulation, and thus averaged over all $z=0$ host halos within the mass range $10^{12}<M<10^{13.7}$ $h^{-1} M_{\Sun}$ to characterize a particular cosmology. A comparison of the averaged distribution is seen in the top panel of Figure~\ref{fig:tform_accrete}. Note that the distribution is computed as a function of look-back time for the given cosmology, not redshift. A small effect is seen demonstrating that in cosmologies with higher $\sigma_8$, $z=0$ subhalos tend to be older. The mean age of subhalos in the $\sigma_8 = 0.9, 0.8,$ and $0.7$ simulations are $11.1, 11.0,$ and $10.6$ billion years respectively. Higher $\sigma_8$ indicates that halos of the same mass should collapse sooner, so this result is expected for host halos. For subhalos, the age distribution can change due to their destruction over time, but evidently this effect is not enough to erase the difference in ages between lasting $z=0$ subhalo populations in different cosmologies.

\subsubsection{Accretion Time of $z=0$ Subhalos}
\label{sec:Accretion}
In addition to formation time, the same analysis is applied to subhalo accretion time. The probability density distribution of subhalo accretion times is similar for all host halos and thus averaged to reduce scatter. The accretion time is defined by when the center of an eventual $z=0$ subhalo crosses its host's $r_{200}$, defined at the time of crossing, and remains inside. This characterization is relevant in that it roughly indicates the time when ram pressure stripping and tidal stripping become important mechanisms of gas loss, which suppresses star formation \citep{Mayer06, Nickerson11}. The accretion time probability density distribution is shown in the bottom panel of Figure~\ref{fig:tform_accrete}. Unlike formation time, no difference inconsistent with scatter is observed between cosmologies. This indicates that subhalos are destroyed as a function of when they were accreted in a process that is independent of small variations in cosmology.

\subsubsection{Maximal Mass}
Finally, we study the averaged mass function per unit $z=0$ host halo mass for the maximal masses of eventual $z=0$ subhalos. As in Figure~\ref{fig:RS_SHMF}, this mass function is scaled to a host of mass $0.84 \times 10^{12}$ $h^{-1}\mathrm{M_\Sun}$. The total stellar mass/luminosity is related to the maximum mass obtained by the subhalo. Since the more concentrated luminous matter is less easily stripped than dark matter, the maximal mass is a much better indicator of stellar mass than the $z=0$ mass of a subhalo \citep{Springel01, Gao04, Guo13}.
 
\begin{figure}[!h]
\includegraphics[width=0.48\textwidth]{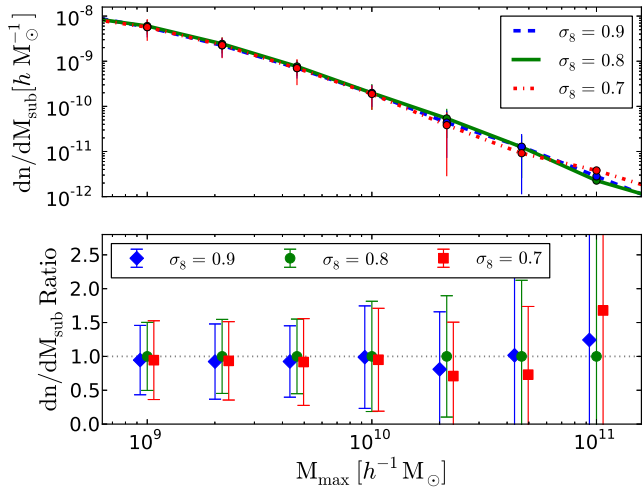}
\caption{\emph{Top Panel:} Average mass function of $M_{max}$ for subhalos that survive until $z=0$, normalized by $z=0$ host halo mass, and scaled to a host of $0.84 \times 10^{12}$ $h^{-1}\mathrm{M_\Sun}$ for cosmologies with different $\sigma_8$. $1\: \sigma$ halo-to-halo variation bars are shown. \emph{Bottom Panel:} Ratio of the subhalo $M_{max}$ function for each cosmology to the fiducial \textit{WMAP}7 cosmology. Data points for the non \textit{WMAP}7 cosmologies are shifted slightly left and right of their true values to help distinguish them. The subhalo $M_{max}$ function is found to not vary with $\sigma_8$. Large variation on a halo-to-halo basis is seen in the $1\: \sigma$ variation bars.}
\label{fig:maxmass}
\end{figure}

Figure~\ref{fig:maxmass} compares the $M_{max}$ mass function for two different cosmologies, showing no dependence on $\sigma_8$. This suggests varying $\sigma_8$ has no effect on the stellar mass/luminosity functions.

\subsection{Properties of Matched Subhalos}
\label{sec:matched}
\begin{figure}[!h]
\includegraphics[width=0.48\textwidth]{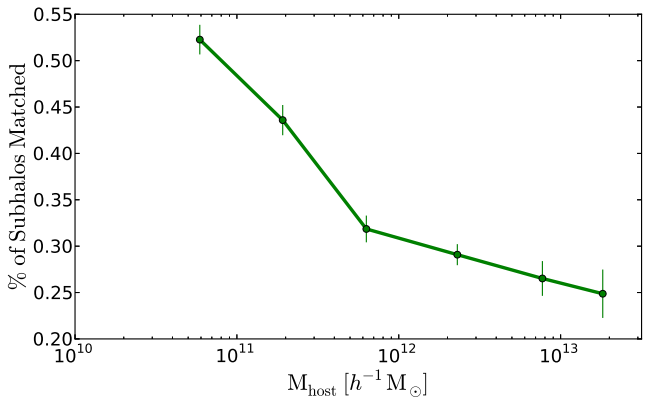}
\caption{Fraction of subhalos in a matched pair of hosts between simulations of different cosmology that are successfully matched to each other as a function of their halo mass. Stronger dynamical processes in the larger halos inhibit subhalo matches more than in smaller host halos. 1$\sigma$ error bars also shown.}
\label{fig:matchedfraction}
\end{figure}

In order to ascertain the difference between specific subhalos due to cosmology, host halos are matched to each other by a procedure described in Section~\ref{sec:matching}. About $90\%$ of hosts above $10^{10}$ $h^{-1}\mathrm{M_\Sun}$ are successfully matched. Subhalos within matched hosts are then matched to each other, and their history is compared. The fraction of subhalos from one host successfully matched to subhalos from the other matched host are shown in Fig.~\ref{fig:matchedfraction}. The remaining unmatched subhalos either are the result of an imperfect matching system, or are ones that correspond to subhalos whose counterpart in the other simulation have already merged with the host, or have not yet been accreted into the host, and are thus not identified as subhalos. The following subsections investigate the same averaged distributions as done in Section \ref{sec:mergertree}, except with only the subset of subhalos that have been matched to each other at $z=0$. Additionally, all figures show only the data matched between $\sigma_8 = 0.8$ and $\sigma_8 = 0.9$ boxes for conciseness. All of the same trends are prevalent in the $\sigma_8 = 0.7$ and $\sigma_8 = 0.8$ matches.

\subsubsection{Formation and Accretion Time}
\label{sec:MatchedFormationAndAccretion}
\begin{figure}[!h]
\includegraphics[width=0.48\textwidth]{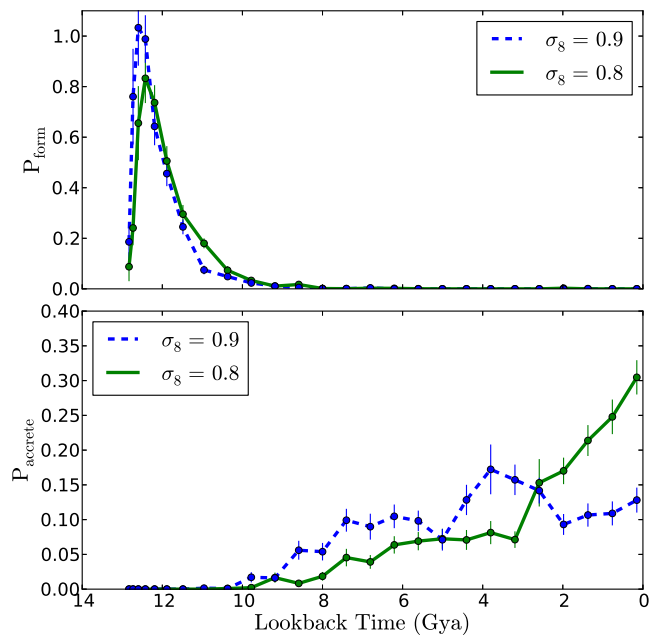}
\caption{\emph{Top Panel}: Average formation time probability density distribution of subhalos matched between $\sigma_8 = 0.9$ and $\sigma_8 = 0.8$ simulations. Matched subhalos in the higher $\sigma_8$ simulation form earlier on average, just as in the general case. \emph{Bottom Panel}: Average accretion time probability density distribution of subhalos matched between $\sigma_8 = 0.9$ and $\sigma_8 = 0.8$ simulations. Unlike the general case, matched subhalos in the higher $\sigma_8$ simulation are accreted significantly earlier. Error bars show $1\: \sigma$ uncertainty of each data point.}
\label{fig:matchedtime}
\end{figure}

Figure~\ref{fig:matchedtime} shows the probability density distributions of matched subhalos for subhalo formation time and accretion time for two different cosmologies. As expected, matched subhalos from the higher $M^*$ simulation, $\sigma_8 = 0.9$, form earlier on average as seen in the top panel of the figure. The difference in the mean ages of the matched subhalos between $\sigma_8 = 0.9$ and $\sigma_8 = 0.8$ is $0.2$ billion years, slightly greater than the $0.1$ billion year difference found in the general sample.

In addition to forming earlier, the matched subhalos in the higher $\sigma_8$ simulation are accreted earlier. This is seen in the bottom panel of Figure~\ref{fig:matchedtime}. Comparison with Figure~\ref{fig:tform_accrete} shows that the matching process preferentially selects subhalos in the $\sigma_8 = 0.9$ simulation that were accreted $\sim 2$ billion years earlier than the typical subhalo.

Thus, in a direct subhalo to subhalo comparison, varying cosmological parameters do have significant effects on the life of a subhalo, even though there are small or no differences found when averaging over all subhalos. Reconciliation of Figure~\ref{fig:matchedtime} and Figure~\ref{fig:tform_accrete} indicates that the unmatched subhalos in the $\sigma_8 = 0.9$ simulation must be ones that were very recently accreted. Following the trend of earlier formation and earlier accretion with higher $\sigma_8$, the counterparts of these unmatched subhalos must be ones that have not yet accreted in the $\sigma_8 = 0.8$ simulation. Similarly, the unmatched subhalos in the $\sigma_8 = 0.8$ simulation are ones that are close to the end of their life. Their counterparts in the $\sigma_8 = 0.9$ simulation are ones that were already tidally destroyed. Therefore, the majority of unmatched subhalos are not stragglers from a faulty matching scheme, but rather the necessary result of different cosmologies leading to a shift in the formation, accretion, and thus destruction of subhalos. Evidently, when studying the accretion times of all subhalos in a host, this accretion and destruction process erases hints of the cosmology the subhalos formed in.

\subsubsection{Maximal Mass and $z=0$ Mass}
\begin{figure}[!h]
\includegraphics[width=0.48\textwidth]{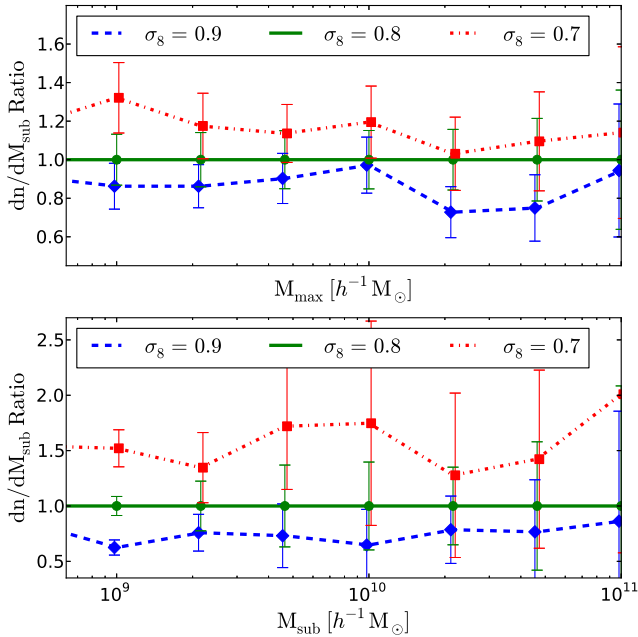}
\caption{\emph{Top Panel}: Ratio of the average mass function of $M_{max}$ normalized by the $z=0$ host halo mass for matched subhalos that survive until $z=0$ to the $M_{max}$ function of the fiducial \textit{WMAP}7 simulation for cosmologies with varied $\sigma_8$. Higher $\sigma_8$ results in lower peak masses in each pair of cosmologies tested. Error bars show $1\: \sigma$ uncertainty of each data point. See Fig.~\ref{fig:maxmass} for the true $M_{max}$ function.
\emph{Bottom Panel}: Ratio of the average SHMF normalized to host halo mass for matched subhalos to the SHMF of the fiducial \textit{WMAP}7 simulation for cosmologies with varied $\sigma_8$. Unlike the general SHMF, the matched SHMF is significantly affected by cosmology. Subhalos in higher $\sigma_8$ simulations end up less massive. Error bars show $1\: \sigma$ uncertainty of each data point. See Fig.~\ref{fig:RS_SHMF} for the true SHMF.}
\label{fig:matchedmass}
\end{figure}

Comparison of the mass function of the maximal mass of matched subhalos, as seen in the top panel of Figure~\ref{fig:matchedmass}, shows a weak dependence on cosmology. Subhalos in the $\sigma_8 = 0.9$ simulation have their life cycle shifted earlier, forming and accreting earlier. The earlier accretion leads to less mass at its peak. Likewise subhalos in the $\sigma_8 = 0.7$ simulation peak at a higher mass than in the $\sigma_8 = 0.8$ simulation. Knowing that such subhalos in the higher $\sigma_8$ simulations accrete earlier, one can expect them to reach $z=0$ at an even smaller mass due to more time spent being tidally stripped. This is confirmed in the mass function displayed in the bottom panel of Figure~\ref{fig:matchedmass}. Matched subhalos are on average $25\%$ less massive in the $\sigma_8 = 0.9$ simulation than the $\sigma_8 = 0.8$ simulation, and $50\%$ more massive in the $\sigma_8 = 0.7$ simulation than the $\sigma_8 = 0.8$ simulation.

\subsubsection{Spatial Distribution}
\begin{figure}[!h]
\includegraphics[width=0.48\textwidth]{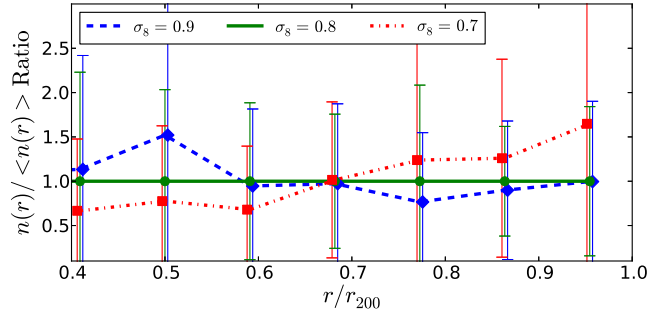}
\caption{The averaged subhalo number density for matched subhalos as a function of $r/r_{vir}$ normalized to the mean matched subhalo number density for cosmologies with varied $\sigma_8$ divided by the same function for the fiducial \textit{WMAP}7 simulation. This demonstrates the trend that matched subhalos in higher $\sigma_8$ cosmologies are closer to the halo center on average. Subhalos are $3.5\%$ closer to the halo center in the $\sigma_8 = 0.9$ simulation than in the $\sigma_8 = 0.8$ simulation on average. Variation bars show $1\: \sigma$ halo-to-halo variation for each data point.}
\label{fig:matchednumdens}
\end{figure}

The spatial distribution of matched subhalos, characterized by number density as a function of $r/r_{200}$ as in Figure~\ref{fig:numdensprofile}, is shown in Figure~\ref{fig:matchednumdens}. The local density is normalized by $<\!\!n\!\!>$, the average number density of matched subhalos within $r_{200}$. Relative to the $\sigma_8 = 0.8$ simulation, the matched subhalos within the $\sigma_8 = 0.9$ simulation are closer to the center of their host: the number density is higher below $r/r_{200} = 0.5$ and lower above $r/r_{200} = 0.8$. The $\sigma_8 = 0.9$ subhalos which accreted earlier on average have more time for dynamical friction effects to slow down their orbits and thus drag them closer to the halo center. The net average change of position is $3.5\%$ closer. This result helps further confirm the conclusions made at the end of Section~\ref{sec:MatchedFormationAndAccretion} that matched subhalos in the $\sigma_8 = 0.9$ simulation are closer to the end of their life and may be destroyed earlier, leaving their counterparts in the $\sigma_8 = 0.8$ simulation matchless.

\section{Summary of Subhalo Property Changes}
\label{sec:summary}
In this investigation, we found that the way in which $\sigma_8, n_s$ and $\Omega_m$ affect subhalo properties can be more concisely summarized in terms of changes in $M^*$, which is positively correlated to all three parameters. The following list details how subhalo properties change with higher $M^*$ cosmologies for all properties that vary with cosmological parameters:
\begin{itemize}
\item Subhalo $V_{max}$ function as a function of host halo mass is greater.
\item Subhalo concentration is greater.
\item Subhalo formation time is earlier on average and on a matched subhalo to subhalo basis.
\item Subhalo accretion time is earlier on a matched subhalo to subhalo basis.
\item Subhalo peak mass is smaller on a matched subhalo to subhalo basis.
\item Subhalo mass at $z=0$ is smaller on a matched subhalo to subhalo basis.
\item Subhalo radial distribution (number density as a function of $r/r_{200}$) is shifted toward the host's center on a matched subhalo to subhalo basis.
\end{itemize}
Individual subhalos in cosmologies with higher $M^*$ are shown to have formed earlier, accreted into their eventual host earlier, both in agreement with \cite{Zentner03, Bosch05}, and as a result of spending more time within their host, lost more mass due to tidal stripping and moved closer to the center of the host due to dynamical friction. This leads to subhalos in higher $M^*$ cosmologies having, on average, less mass at $z=0$. This is opposite of the trend for host halo masses, where large hosts have more mass in higher $M^*$ cosmologies.

Including subhalo abundances, the averaged subhalo properties that remain unchanged under variations in cosmological parameters are:
\begin{itemize}
\item subhalo mass function as a function of host halo mass,
\item subhalo abundance above $V_{max} = 30 \ \mathrm{km \ s^{-1}}$ as a function of host halo mass,
\item subhalo mass fraction as a function of host halo mass,
\item radial distribution of subhalos (number density as a function of $r/r_{200}$),
\item subhalo accretion time probability distribution function, and
\item distribution of maximal mass attained by subhalos before they enter their host's tidal field.
\end{itemize}

\section{Discussion and Conclusion}
\label{sec:conclusion}
Changes in the cosmological parameters $\sigma_8, n_s,$ and $\Omega_m$ affect the abundance of dark matter halos in the universe. Larger values of $n_s$ and $\Omega_m$ lead to a larger abundance of halos. Larger values of $\sigma_8$ lead to a higher abundance of halos above the current characteristic mass scale of collapse, $M^*$, and a lower abundance below that threshold. These differences, however, as tested by a suite of cosmological dark matter only simulations, are erased when considering local subhalo populations (all subhalos within a given host halo) as a function of their host halo mass. For any given host halo mass, the average subhalo mass function is independent of small variations in cosmological parameters on the order of the changes in the best estimate values of \textit{WMAP} and \textit{Planck} \citep{Planck13}. Thus, the expected abundance of dark matter subhalos by mass in the Milky Way, for example, should not change with the recent revision of parameters by \textit{Planck}.

Subhalo concentrations, $V_{max}$ functions, and formation times, on the other hand, retain the same trends in cosmological dependence as the host halos. In cosmologies where objects tend to collapse sooner (higher values of $\sigma_8$, $n_s$, and $\Omega_m$), both host halos and subhalos have systematically more concentrated cores for all mass ranges considered and form earlier. More concentrated cores lead to subhalos with higher $V_{max}$ relative to less concentrated halos of the same mass. Other cosmology induced differences arise on a subhalo to subhalo basis. Since all simulations used the same initial random white noise field, it was possible to identify halos and subhalos which arose from the same density fluctuation. A comparison of these ``matched'' subhalos indicates that cosmological parameters do in fact have a significant effect on individual halos, even ones that become subhalos. In the higher $\sigma_8$ simulation, matched subhalos formed earlier, accreted earlier, are located closer to the host halo center, are more concentrated, and are smaller in mass.

Clearly, there is a disparity of results between averaging properties of all subhalos within a host and averaging only the subset of matched subhalos. This indicates that the mechanisms of subhalo accretion, mass loss, and ultimately destruction lead to a cosmology independent steady state distribution of subhalos within a host. While subhalos in the higher $M^*$ simulations are accreted earlier on average, there is a steady rate of subhalo accretion that is the same in all cosmologies at low redshift for hosts of the same size. In addition, even though the overall mass function of halos changes, the mass function of subhalos at accretion for a given host halo does not change. Since subhalo mass and angular momentum loss is governed by cosmology independent physics, it follows that the resulting size, position, number, and accretion time of subhalos will be unaffected by cosmology. In contrast, the formation time and concentration, and thus $V_{max}$, of subhalos at accretion do vary with cosmology. These differences are not erased subhalo-host interactions.

The results of this study indicate that in simulations with pre-\textit{Planck} cosmological parameters, characterizations of subhalo mass function, mass fraction, spatial distribution, accretion time, and peak mass at infall are all still correct. This indicates that expensive high resolution simulations remain valid when used to study subhalos as long as the properties are appropriately normalized to the mass of the host. There are a few exceptions for when subhalo concentration matters, such as in computing a dark matter annihilation rate estimate from substructure, or when subhalo formation time (defined by when halos first reach a threshold mass) is important, or to a lesser extent when the subhalo $V_{max}$ function matters. In such cases a simple way to estimate how these properties change is to compute $M^*$ for two cosmologies and know that in higher $M^*$ cosmologies subhalos will be more concentrated and have formed earlier. For example, between \textit{WMAP}7 and \textit{Planck}, $M^*$ increased by an amount similar to the effect of an increase of $\sigma_8$ of $0.07$. The results also indicate that variations in cosmological parameters should have negligible effect on the magnitude of the missing satellite problem, and some affect on the too big to fail problem. Decreases in predicted subhalo $V_{max}$ and concentration, which could come from lower values of $n_s$, $\sigma_8$, and $\Omega_m$, both serve to alleviate the too big to fail problem. \cite{Polisensky13} confirms this for the cases of $n_s$, $\sigma_8$. Still, based on the study of subhalo properties over many host halos, the mass of the host halo and variations from halo to halo are much more important in controlling the distribution of subhalo properties within a galaxy than cosmological parameters.
\acknowledgements
We thank Peter Behroozi for help with using {\sc{rockstar}} and Volker Springel for giving access to {\sc{gadget3}}. G.A.D. acknowledges support from an NSF Graduate Research Fellowship under Grant No. 1122374. \\

\end{document}